\documentclass[prb,twocolumn ]{revtex4}
\usepackage[dvips]{graphicx}

\begin{document}

\bibliographystyle{apsrev}

\title{All--optical generation and detection of sub--picosecond ac spin current pulses in GaAs}
\author{Brian A. Ruzicka, Karl Higley, Lalani K. Werake, and Hui Zhao\footnote{Corresponding author, huizhao@ku.edu}}

\affiliation{Department of Physics and Astronomy, The University of Kansas, Lawrence, Kansas 66045}

\begin{abstract}
Sub--picosecond ac spin current pulses are generated optically in GaAs bulk and quantum wells at room temperature and 90~K through quantum interference between one--photon and two--photon absorptions driven by two phase--locked ultrafast laser pulses that are both circularly polarized. The dynamics of the current pulses are detected optically by monitoring in real time and real space nanoscale motion of electrons with high--resolution pump--probe techniques. The spin polarization of the currents is $0.6\pm 0.1$, with peak current densities on the order of $10^2~\mathrm{A/m^2}$.
\end{abstract}

\maketitle

Generation, manipulation and detection of spin currents in semiconductors are the fundamental aims of spintronics.\cite{s2941488,rmp76323} Although it is possible to generate pure spin currents that are not accompanied by any charge currents through a spin Hall effect\cite{s3061910,l94047204,l96246601} or some optical techniques\cite{l90136603,l90216601,b72201302}, in most cases, spin currents are carried by charge currents. Compared to pure spin currents, these spin--polarized charge currents are easier to produce and manipulate, and therefore have been used in most spintronic designs.

In the past, spin--polarized charge currents have been generated by dragging optically excited spin--polarized carriers by an electric field,\cite{apl731580,n397139} or though contact with magnetic materials.\cite{n401572,s3092191,n447295} In these configurations, the currents are dc and, in most cases, steady state. However, for high speed spintronic applications, short spin current pulses are desirable. Furthermore, similar to charge--based electronics, ac spin currents may have advantages over dc spin currents in some configurations. Although there have been theoretical proposals on electrical\cite{b68233307} and optical\cite{b73201302} generations of ac spin currents, we are not aware of any experimental demonstrations to date. In addition, attempting to generate ultrashort spin current pulses is rare.

Optical techniques have the potential to produce ultrashort spin current pulses since ultrafast laser pulses can be readily produced. Previously, optical injections of spin--polarized charge currents have been demonstrated through quantum interference in bulk GaAs\cite{jap914382} and by spin photo--galvanic effect in several structures including GaAs quantum wells (QWs)\cite{apl773146,b68035319,njp9349}, InAs QWs\cite{l864358}, Si/Ge QWs\cite{apl91252102}, and AlGaN/GaN superlattices\cite{b75085327,apl90041909,apl91071912,apl91071920}. These spin currents are generated pure optically, without applying any external electric fields. However, in each of these studies currents were detected not by optical techniques, but by measuring steady--state voltage\cite{jap914382,apl773146,b68035319,njp9349,l864358,apl91252102,b75085327,apl90041909,apl91071912,apl91071920} caused only by the charge component of the currents. Therefore, not only has the degree of spin polarization of these photogenerated currents not been measured, but
strictly speaking, even the existence of a spin component of these currents has not been experimentally demonstrated, since spin--polarized carriers do not necessarily carry a spin--polarized current. More importantly, the steady--state electric detection technique cannot measure the temporal evolution of the currents; therefore the dynamics of these photogenerated currents have not been studied.

Here we demonstrate all--optical generation and detection of sub--ps ac spin current pulses in GaAs. Spin--polarized charge currents are injected all optically by utilizing quantum interference between one--photon and two--photon absorptions driven by two phase--locked and circularly polarized ultrafast laser pulses. In contrast to previous steady--state measurements\cite{jap914382,apl773146,b68035319,njp9349,l864358,apl91252102,b75085327,apl90041909,apl91071912,apl91071920}, we time resolve the dynamics of the currents by monitoring in real space and real time nanoscale carrier transport by using spatially and temporally resolved pump--probe techniques. This allows us to demonstrate that the photogenerated currents are ac in nature and pulsed with a pulsewidth shorter than 1 ps. Furthermore, by using optical detection techniques, we simultaneously monitor the charge and spin components of the currents and therefore are able to measure the spin polarization degree of the currents of 0.6$\pm 0.1$ in both bulk and QW samples, which agrees reasonably with theoretical predictions.\cite{l855432,b68165348} The peak current densities achieved are on the order of $10^2~\mathrm{A/m^2}$.

In the experiments, we simultaneously illuminate GaAs samples with two right--hand circularly polarized and phase--locked laser pulses with angular frequencies $\omega$ and 2$\omega$, as shown in Fig.~1a. Under the condition of $\hbar \omega < E_{\mathrm{g}} <2\hbar \omega$, where $E_{\mathrm{g}}$ is the bandgap of GaAs, the $\omega$ and 2$\omega$ pulses excite electrons from the valence band to the conduction band via two--photon and one--photon absorptions, respectively (inset of Fig.~1c). In this configuration, spin--polarized charge currents are injected through quantum interference between the two transition pathways driven by the two pulses.\cite{l855432,b68165348} By arranging the phases of the two pulses, the transition amplitudes interfere constructively at some k states, but destructively at opposite k states, resulting in an asymmetric distribution of carriers in k--space, as shown in Fig. 1c (inset). Spin--polarized electrons are injected in the conduction band with an average velocity that can be phenomenologically written as $\langle v \rangle =v_{0} \eta [\mathrm{sin}(\Delta \phi) \hat{x}+\mathrm{cos}(\Delta \phi) \hat{y}]$, where $v_{0}$ is the speed of each electron and $\eta$ describes current injection efficiency.\cite{l855432,b68165348} The relative phase between the two pulses, $\Delta \phi \equiv \phi_{2\omega}-2\phi_{\omega}$, controls the direction of the velocity in the x--y plane. In this work, we inject and detect currents along $\hat{x}$ by choosing $\Delta \phi=\pi /2$. Since both heavy--hole and light--hole transitions are excited in this configuration, the spin polarization of {\it electrons} is expected to be about 0.5 in both bulk and QW structures, according to the well--established spin selection rules.\cite{opticalorientation} However, theories predict slightly larger spin polarizations of the {\it currents} of 0.57 in bulk\cite{l855432} and QW structures\cite{b68165348}.

\begin{figure}
\includegraphics[width=7.5cm]{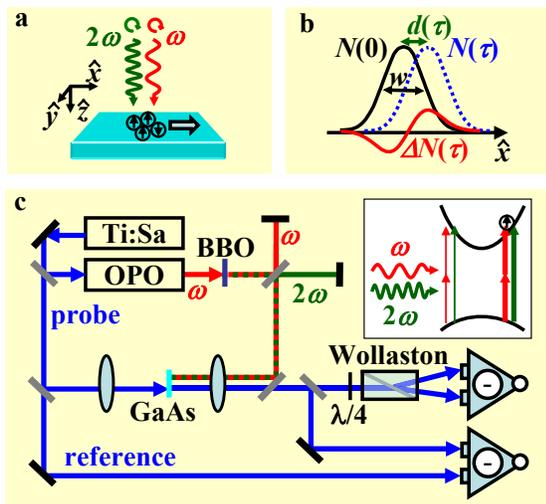}
\caption{(Color online) (a) Experimental configuration showing injection of a spin--polarized charge current by illuminating a GaAs sample with two harmonically related and circularly polarized pulses; (b) Electron density profiles upon injection (solid Gaussian curve) and $\tau$ later (dotted Gaussian curve) and the difference between the two profiles, electron accumulation (solid derivative--like curve); (c) Experimental set--up and excitation scheme (inset).}
\end{figure}

The dynamics of the currents are studied by monitoring the charge and spin transport along $\hat{x}$. As shown in Fig.~1b, electrons are injected in the conduction band with a density profile $N(0)$ (solid Gaussian curve) and an average velocity along $+\hat{x}$ direction when $\Delta \phi = \pi /2$. After a time period of $\tau$, the profile moves to a new position $N(\tau)$ (dotted Gaussian curve). Under typical conditions, the transport length $d$ is much smaller than the width $w$. Therefore, the difference between the two profiles, $\Delta N \equiv N(\tau)-N(0)$, i.e., electron accumulation due to transport, has a derivative--like spatial profile. It is straightforward to show that the height of this profile is proportional to $d$, i.e., $\Delta N_{\mathrm{MAX}}=1.4 (N_{\mathrm{MAX}}/w)d$.\cite{b72201302} Hence, by measuring the temporal evolutions of $\Delta N_{\mathrm{MAX}}$, $N_{\mathrm{MAX}}$ and $w$, we can obtain the temporal evolution of $d$.

The experiments are performed on two GaAs samples at two temperatures: a 400--nm thick bulk sample at room temperature and a multiple QW sample at 90~K, which is composed of 10 periods of 14--nm GaAs layers sandwiched by 14--nm AlGaAs barriers. In the following, we will first discuss procedures and results of the experiment on the QW sample at 90~K, and then briefly discuss the results of the bulk sample measurement at room temperature.

To inject currents in the QW sample, the $\omega$ pulse with a central wavelength of 1500~nm and a pulse width of 100~fs is obtained from an optical parametric oscillator pumped by a Ti:sapphire laser at 80~MHz (Fig.~1c). The $2\omega$ pulse is obtained by frequency doubling the $\omega$ pulse with a beta barium borate (BBO) crystal. The two pulses are sent through a dichroic interferometer for phase control, and then combined and focused to the sample. By using a series of quarter--wave plates and polarizers in the interferometer (not shown), the polarizations of both pulses are set to be right hand circular with purities better than 97\% on the sample.  The $2\omega$ pulse is tightly focused to a spot size of 1.8~$\mu$m full width at half maximum (FWHM) with a peak fluence of 5 $\mu \mathrm{J/cm^{2}}$. It excites spin--polarized carriers by interband one--photon absorption (inset of Fig.~1c) with a peak areal density of $2 \times \mathrm{10^{12}/cm^{2}}$. The fluence and beam size of the $\omega$ pulse are set to produce the same peak density and size of carrier profile through two--photon absorption.

Electron densities are measured by focusing a linearly polarized 100~fs probe pulse obtained from the Ti:sapphire laser on the sample with a spot size of 1.8~$\mu$m FWHM, as shown in Fig.~1c. A differential transmission $\Delta T(N)/T_0 \equiv [T(N)-T_0)]/T_0$, i.e., the normalized difference between transmissions with [$T(N)$] and without [$T_0$] carrier's presence, is measured by reflecting a portion of the transmitted probe pulse to a photodiode of a balanced detector (lower part of Fig.~1c) connected to a lock--in amplifier referenced to a chopper in the pump beam (not shown). A reference pulse is sent to the other photodiode of that balanced detector in order to suppress laser intensity noise.\cite{apl92112104} The probe pulse is tuned to heavy--hole excitonic resonance (808~nm) to selectively probe electrons based on well--established excitonic absorption saturation caused by free carriers.\cite{b326601} Compared to electrons, holes make smaller contribution to $\Delta T(N)/T_0$ since heavy holes have a larger effective mass\cite{b326601} and light holes do not directly saturate the heavy--hole excitonic absorption. Therefore, for simplicity, we take $\Delta T(N)/T_0$ as a measurement of electron density only. Furthermore, we verify by measuring $\Delta T(N)/T_0$ as a function of pump pulse fluence that, for the carrier densities used in this study, $\Delta T(N)/T_0 \propto N$.

Electron accumulation $\Delta N$ is detected by measuring a phase--dependent differential transmission $\Delta T(\Delta \phi)/T(\Delta \phi =0) \equiv [T(\Delta \phi)-T(\Delta \phi =0)]/T(\Delta \phi =0)$, i.e., the normalized difference between transmissions with [$\Delta T(\Delta \phi)$] and without [$\Delta T(\Delta \phi =0)$] current injection. This is done by modulating $\Delta \phi$ by mechanically dithering the length of one arm of the interferometer at 37~Hz using a piezoelectric transducer and then measuring the output of the balanced detector with a lock--in amplifier that is slaved to the modulation frequency.\cite{l96246601}

Spin density $S \equiv N_{\uparrow}-N_{\downarrow}$, where $N_{\uparrow}$ ($N_{\downarrow}$) is the density of electrons with spin along + (-)$\hat{z}$, is measured by analyzing carrier--induced circular dichroism, i.e., the absorption difference of right-- and left--hand circularly polarized probe pulses in the presence of spin--polarized carriers.\cite{b72201302,l96246601} The linearly--polarized probe pulse used in the experiment is composed of two circular components. Due to spin--selection rules, each component preferentially senses one spin system.\cite{opticalorientation} By using a quarter--waveplate ($\lambda /4$) and a Wollaston prism, we send the two components to two photodiodes of another balanced detector (Fig.~1c). The output of the balanced detector is proportional to the difference between the differential transmissions of the two circular components, $(\Delta T^+ -\Delta T^-)/T_0$, which is proportional to $S$.\cite{b72201302,l96246601}
With this configuration, when $\Delta \phi$ is modulated to measure $\Delta N$, the upper balanced detector (Fig.~1c) simultaneously outputs the spin accumulation due to spin transport, $\Delta S \equiv S(\tau)-S(0)$.
Finally, the measured circular dichroism is related to spin density by using a calibration process based on the well--established fact that interband transition induced by a circularly polarized pump pulse produces a spin--polarization $S/N = 0.5$.\cite{opticalorientation}

\begin{figure}
 \includegraphics[width=8.5cm]{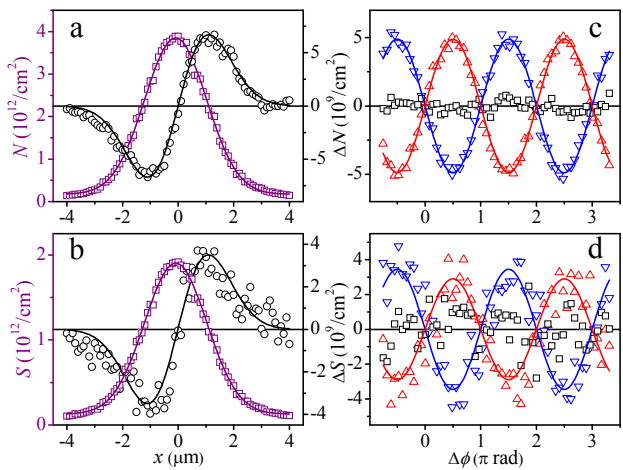}
 \caption{(Color online) Profiles of electron density (squares in a), electron accumulation (circles in a), spin density (squares in b) and spin accumulation (circles in b) measured with a probe delay of 0.3 ps and $\Delta \phi=\pi /2$ on the quantum--well sample at 90~K. From the two profiles in panel a, a transport length of 5~nm is deduced using the procedure illustrated in Fig. 1b. Panel c (d) shows electron (spin) accumulation measured at a probe position of $x=+1.7~\mu$m (up--triangles), $-1.7~\mu$m (down--triangles) and 0 (squares), respectively, when $\Delta \phi$ is varied.}
\end{figure}

Figure 2 summarizes measurements on the QW sample at 90~K performed with a fixed probe delay of 0.3 ps. The spatial profiles of $N$ (a, squares), $\Delta N$ (a, circles), $S$ (b, squares) and $\Delta S$ (b, circles) are measured by scanning the probe spot along $\hat{x}$ with $\Delta \phi=\pi /2$. The Gaussian profiles of $N$ and $S$ are consistent with the shape and size of the laser spots. The derivative--like $\Delta N$ profile shows that electrons accumulate and deplete along $\hat{x}$, indicating that the electron density profile has moved along $+\hat{x}$. From these profiles, we obtain the transport length $d = 5~\mathrm{nm}$ by using the previously mentioned formula. Spin transport is also evident from the derivative--like $\Delta S$ profile. We therefore demonstrate that the photogenerated currents are indeed spin--polarized, since for a pure charge current, the accumulated electrons should be spin--unpolarized and $\Delta S$ should be zero. Quantitatively, we find that the spin polarization of the accumulated electrons due to the current $\Delta S / \Delta N \approx 0.6$. This indicates the spin--polarization of the current is also about 0.6.

Panels c and d of Fig.~2 demonstrate phase control of the current injection. Up--triangles in c show $\Delta N$ as a function of $\Delta \phi$ measured at $x=+1.7~\mu$m. The observed sinusoidal variation is consistent with the sinusoidal $\Delta \phi$ dependence of the injected average velocity. The sinusoidal $\Delta \phi$ dependence is also observed at the other side of the profile with $x=-1.7~\mu$m (down--triangles). The two curves are $\pi$ out--of--phase. Furthermore, measurement performed at $x=0$ yields no signal above noise level (squares). All of these are consistent with the derivative--like profile of $\Delta N$ seen in panel a. Similar results are also obtained for the spin accumulation $\Delta S$, as shown in panel d.

\begin{figure}
 \includegraphics[width=8cm]{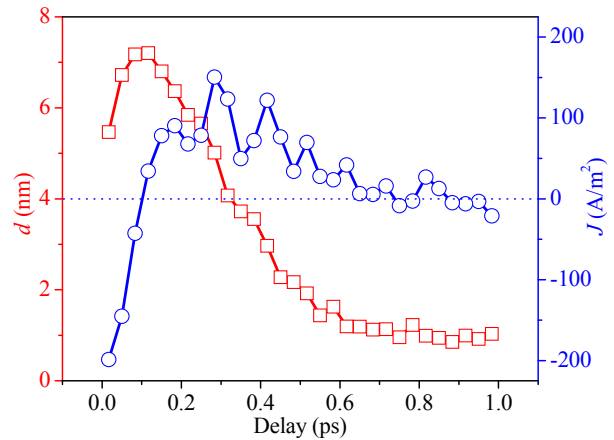}
 \caption{(Color online) Temporal evolutions of the transport length (squares) and current density (circles) on the quantum--well sample at 90~K obtained by repeating measurements summarized in Fig.~2 with different probe delays.}
\end{figure}

Optical detection techniques based on ultrafast lasers provide us enough high temporal resolution to time resolve the current dynamics. The procedure summarized in Fig.~2 is used to measure $d$ as a function of probe delay. The results are shown as the squares in Fig.~3. Quantitative modelings of the current dynamics are beyond the scope of this experimental work, however we provide the following qualitative explanation. By using quantum interference, spin--polarized electrons are injected with an average velocity along $+\hat{x}$. Therefore, upon injection, the electrons move along $+\hat{x}$. The same quantum interference process also injects holes with opposite momentum, according to crystal momentum conservation. Therefore, holes move along $-\hat{x}$. Since heavy holes have a larger effective mass, they move with a smaller average velocity. Once the electrons and holes separate, a space charge field develops, slowing down and eventually stopping the motions of electrons and holes. Then, the space charge field becomes a driving force to pull the electrons and holes back to a common location. Since during the whole process, strong phonon and inter--carrier scatterings exist, this oscillator--like system is strongly damped, and therefore multiple oscillations are not observed. The dynamics exist only for less than 1 ps.
Apparently, although the holes only make weak contributions to the differential transmission of the probe, they do play important roles in determining the current dynamics.

The squares of Fig.~3 show the temporal evolution of the average position of electrons. A time derivative of that curve gives the temporal evolution of the average velocity, and therefore the charge current density, as shown with the circles in Fig.~3. Despite large uncertainties of the data due to a poor signal--to--noise ratio, the ac and ultrashort nature of the current is obvious, as one would infer from the temporal evolution of $d$. The current starts with highest and negative density due to the instantaneous optical injection. It decays with time, then changes to positive, and eventually decays to zero within 1 ps. We therefore demonstrate generation of ac sub--ps spin current pulses. Due to the ultrashort pulsewidth, the ac current is single cycle.

We emphasis that, although Fig.~3 only shows the charge component of the current, the spin component is simultaneously monitored in the experiment, with similar temporal behaviors observed. When taking the ratio, no temporal variation of the $\Delta S / \Delta N$ is observed on the time scale of 1~ps. This is consistent with the long spin relaxation time of about 100~ps that is measured separately by monitoring decay of $S/N$ on longer time scales. By averaging the data, we obtain the spin--polarization of the accumulated electrons, and thus the spin--polarization of the current, to be 0.6$\pm 0.1$. This value is reasonably consistent with earlier theoretical prediction of 0.57.\cite{l855432,b68165348}

\begin{figure}
 \includegraphics[width=8.5cm]{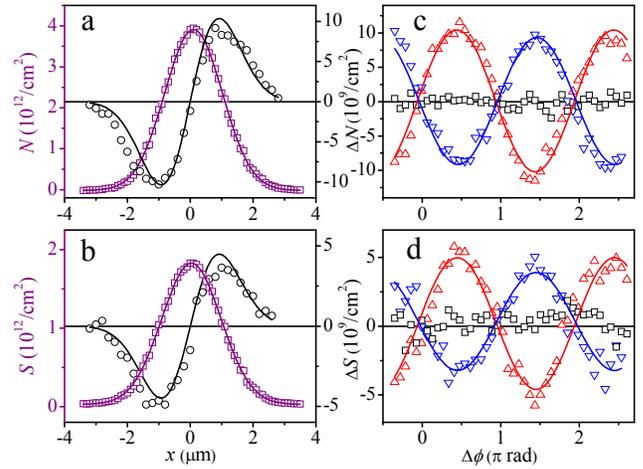}
 \caption{(Color online) Profiles of electron density (squares in a), electron accumulation (circles in a), spin density (squares in b) and spin accumulation (circles in b) measured with a probe delay of 0.3 ps and $\Delta \phi=\pi /2$ on the bulk sample at room temperature. From the two profiles in panel a, a transport length of 3.8~nm is deduced using the procedure illustrated in Fig. 1b. Panel c (d) shows electron (spin) accumulation measured at a probe position of $x=+1.0~\mu$m (up--triangles), $-1.0~\mu$m (down--triangles) and 0 (squares), respectively, when $\Delta \phi$ is varied.}
\end{figure}

\begin{figure}
 \includegraphics[width=8cm]{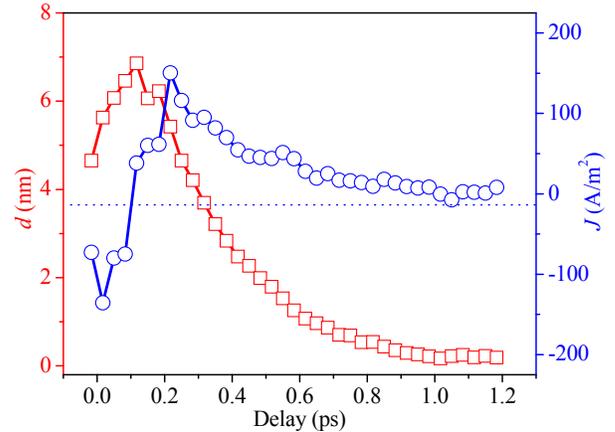}
 \caption{(Color online) Temporal evolutions of the transport length (squares) and current density (circles) on the bulk sample at room temperature obtained by repeating measurements summarized in Fig.~4 with different probe delays.}
\end{figure}

The experiment on the bulk sample at room temperature is carried out under the same excitation conditions. However, a few changes are made in the detection scheme. At room temperature excitonic resonances are thermally broadened, and therefore significantly overlap with band--to--band transitions. It is difficult to selectively probe the heavy--hole excitonic transition, as we did on the QW sample at 90~K for efficiently sensing carriers and their spin. The probe pulse is tuned to a central wavelength of 820~nm, corresponding to an excess energy of 90~meV. This causes the differential transmission signal $\Delta T(N)/T_0$ to drop by a factor of 20 with the same carrier density. Increasing the probe pulse wavelength would improve the probing efficiency, however, the pumping efficiency of the optical parametric oscillator drops severely.

In order to detect the much reduced signal, we improve our detection capability by increasing the modulation frequency. This is achieved by replacing the mechanical modulation device (piezoelectric transducer) used in the low--temperature measurement, which can only operate below 40~Hz, by an electro--optical phase modulator with 2439~Hz modulation frequency. Although the 50--mm thick lithium niobate crystal in the phase modulator dispersively broadens the 2$\omega$ pulse to about 300 fs, resulting in a lower temporal resolution of the study, it reduces the noise level by two orders of magnitude, according to the noise spectra of our laser and detection systems.

In Fig.~4 we show the measurements on the bulk sample at room temperature performed with a probe delay of 0.3~ps in a similar fashion of Fig.~2. The improvement in signal--to--noise ratio is evident in Fig.~4 in comparison with Fig.~2, since signals in the room temperature measurement are about 20 times weaker than at 90~K . Using similar procedures, we deduce a transport length $d = 3.8~\mathrm{nm}$ and spin polarization of the accumulated electrons due to the current $\Delta S / \Delta N \approx 0.6$. Finally, the measurements summarized in Fig.~4 are repeated at different probe delays to temporally resolve the dynamics. The results are shown in Fig.~5.

The dynamics observed in the two experiments are similar. The current injection processes are expected to be similar in bulk and QW structures according to theoretical calculations.\cite{l855432,b68165348} The current dynamics are determined by a number of factors including inter--carrier scatterings, phonon scatterings, and space--charge field. Since phonon absorption rates increase with temperature, one would expect faster momentum relaxation at room temperature, suggesting a smaller maximum transport length. The similar maximum transport lengths observed at 90~K and room temperature indicates that momentum relaxation is likely dominated by inter--carrier scatterings and phonon emissions under these conditions.

In summary, we demonstrate all--optical generation and detection of sub--ps ac spin current pulses in GaAs bulk and QW structures at room temperature and 90~K. The currents and their spin--polarization are detected by spatially and temporally resolving nanoscale motion of electrons with high--resolution pump--probe techniques. The spin polarization of the currents is measured to be 0.6$\pm 0.1$, with a peak current density on the order of $10^2~\mathrm{A/m^2}$.

We acknowledge John Prineas of University of Iowa for providing us with high--quality GaAs samples, and financial support of the General Research Fund of The University of Kansas.


\begin{thebibliography}{30}
\expandafter\ifx\csname natexlab\endcsname\relax\def\natexlab#1{#1}\fi
\expandafter\ifx\csname bibnamefont\endcsname\relax
  \def\bibnamefont#1{#1}\fi
\expandafter\ifx\csname bibfnamefont\endcsname\relax
  \def\bibfnamefont#1{#1}\fi
\expandafter\ifx\csname citenamefont\endcsname\relax
  \def\citenamefont#1{#1}\fi
\expandafter\ifx\csname url\endcsname\relax
  \def\url#1{\texttt{#1}}\fi
\expandafter\ifx\csname urlprefix\endcsname\relax\def\urlprefix{URL }\fi
\providecommand{\bibinfo}[2]{#2}
\providecommand{\eprint}[2][]{\url{#2}}

\bibitem[{\citenamefont{Wolf et~al.}(2001)\citenamefont{Wolf, Awschalom,
  Buhrman, Daughton, von Molnar, Roukes, Chtchelkanova, and Treger}}]{s2941488}
\bibinfo{author}{\bibfnamefont{S.~A.} \bibnamefont{Wolf}},
  \bibinfo{author}{\bibfnamefont{D.~D.} \bibnamefont{Awschalom}},
  \bibinfo{author}{\bibfnamefont{R.~A.} \bibnamefont{Buhrman}},
  \bibinfo{author}{\bibfnamefont{J.~M.} \bibnamefont{Daughton}},
  \bibinfo{author}{\bibfnamefont{S.}~\bibnamefont{von Molnar}},
  \bibinfo{author}{\bibfnamefont{M.~L.} \bibnamefont{Roukes}},
  \bibinfo{author}{\bibfnamefont{A.~Y.} \bibnamefont{Chtchelkanova}},
  \bibnamefont{and} \bibinfo{author}{\bibfnamefont{D.~M.}
  \bibnamefont{Treger}}, \bibinfo{journal}{Science}
  \textbf{\bibinfo{volume}{294}}, \bibinfo{pages}{1488} (\bibinfo{year}{2001}).

\bibitem[{\citenamefont{$\mathrm{\check{Z}uti\acute{c}}$
  et~al.}(2004)\citenamefont{$\mathrm{\check{Z}uti\acute{c}}$, Fabian, and
  Sarma}}]{rmp76323}
\bibinfo{author}{\bibfnamefont{I.}~\bibnamefont{$\mathrm{\check{Z}uti\acute{c}%
}$}}, \bibinfo{author}{\bibfnamefont{J.}~\bibnamefont{Fabian}},
  \bibnamefont{and} \bibinfo{author}{\bibfnamefont{S.~D.} \bibnamefont{Sarma}},
  \bibinfo{journal}{Rev. Mod. Phys.} \textbf{\bibinfo{volume}{76}},
  \bibinfo{pages}{323} (\bibinfo{year}{2004}).

\bibitem[{\citenamefont{Kato et~al.}(2004)\citenamefont{Kato, Myers, Gossard,
  and Awschalom}}]{s3061910}
\bibinfo{author}{\bibfnamefont{Y.~K.} \bibnamefont{Kato}},
  \bibinfo{author}{\bibfnamefont{R.~C.} \bibnamefont{Myers}},
  \bibinfo{author}{\bibfnamefont{A.~C.} \bibnamefont{Gossard}},
  \bibnamefont{and} \bibinfo{author}{\bibfnamefont{D.~D.}
  \bibnamefont{Awschalom}}, \bibinfo{journal}{Science}
  \textbf{\bibinfo{volume}{306}}, \bibinfo{pages}{1910} (\bibinfo{year}{2004}).

\bibitem[{\citenamefont{Wunderlich et~al.}(2005)\citenamefont{Wunderlich,
  Kaestner, Sinova, and Jungwirth}}]{l94047204}
\bibinfo{author}{\bibfnamefont{J.}~\bibnamefont{Wunderlich}},
  \bibinfo{author}{\bibfnamefont{B.}~\bibnamefont{Kaestner}},
  \bibinfo{author}{\bibfnamefont{J.}~\bibnamefont{Sinova}}, \bibnamefont{and}
  \bibinfo{author}{\bibfnamefont{T.}~\bibnamefont{Jungwirth}},
  \bibinfo{journal}{Phys. Rev. Lett.} \textbf{\bibinfo{volume}{94}},
  \bibinfo{pages}{047204} (\bibinfo{year}{2005}).

\bibitem[{\citenamefont{Zhao et~al.}(2006)\citenamefont{Zhao, Loren, van Driel,
  and Smirl}}]{l96246601}
\bibinfo{author}{\bibfnamefont{H.}~\bibnamefont{Zhao}},
  \bibinfo{author}{\bibfnamefont{E.~J.} \bibnamefont{Loren}},
  \bibinfo{author}{\bibfnamefont{H.~M.} \bibnamefont{van Driel}},
  \bibnamefont{and} \bibinfo{author}{\bibfnamefont{A.~L.} \bibnamefont{Smirl}},
  \bibinfo{journal}{Phys. Rev. Lett.} \textbf{\bibinfo{volume}{96}},
  \bibinfo{pages}{246601} (\bibinfo{year}{2006}).

\bibitem[{\citenamefont{Stevens et~al.}(2003)\citenamefont{Stevens, Smirl,
  Bhat, Najmaie, Sipe, and van Driel}}]{l90136603}
\bibinfo{author}{\bibfnamefont{M.~J.} \bibnamefont{Stevens}},
  \bibinfo{author}{\bibfnamefont{A.~L.} \bibnamefont{Smirl}},
  \bibinfo{author}{\bibfnamefont{R.~D.~R.} \bibnamefont{Bhat}},
  \bibinfo{author}{\bibfnamefont{A.}~\bibnamefont{Najmaie}},
  \bibinfo{author}{\bibfnamefont{J.~E.} \bibnamefont{Sipe}}, \bibnamefont{and}
  \bibinfo{author}{\bibfnamefont{H.~M.} \bibnamefont{van Driel}},
  \bibinfo{journal}{Phys. Rev. Lett.} \textbf{\bibinfo{volume}{90}},
  \bibinfo{pages}{136603} (\bibinfo{year}{2003}).

\bibitem[{\citenamefont{H$\mathrm{\ddot{u}}$bner
  et~al.}(2003)\citenamefont{H$\mathrm{\ddot{u}}$bner, R$\mathrm{\ddot{u}}$hle,
  Klude, Hommel, Bhat, Sipe, and van Driel}}]{l90216601}
\bibinfo{author}{\bibfnamefont{J.}~\bibnamefont{H$\mathrm{\ddot{u}}$bner}},
  \bibinfo{author}{\bibfnamefont{W.~W.} \bibnamefont{R$\mathrm{\ddot{u}}$hle}},
  \bibinfo{author}{\bibfnamefont{M.}~\bibnamefont{Klude}},
  \bibinfo{author}{\bibfnamefont{D.}~\bibnamefont{Hommel}},
  \bibinfo{author}{\bibfnamefont{R.~D.~R.} \bibnamefont{Bhat}},
  \bibinfo{author}{\bibfnamefont{J.~E.} \bibnamefont{Sipe}}, \bibnamefont{and}
  \bibinfo{author}{\bibfnamefont{H.~M.} \bibnamefont{van Driel}},
  \bibinfo{journal}{Phys. Rev. Lett.} \textbf{\bibinfo{volume}{90}},
  \bibinfo{pages}{216601} (\bibinfo{year}{2003}).

\bibitem[{\citenamefont{Zhao et~al.}(2005)\citenamefont{Zhao, Pan, Smirl, Bhat,
  Najmaie, Sipe, and van Driel}}]{b72201302}
\bibinfo{author}{\bibfnamefont{H.}~\bibnamefont{Zhao}},
  \bibinfo{author}{\bibfnamefont{X.}~\bibnamefont{Pan}},
  \bibinfo{author}{\bibfnamefont{A.~L.} \bibnamefont{Smirl}},
  \bibinfo{author}{\bibfnamefont{R.~D.~R.} \bibnamefont{Bhat}},
  \bibinfo{author}{\bibfnamefont{A.}~\bibnamefont{Najmaie}},
  \bibinfo{author}{\bibfnamefont{J.~E.} \bibnamefont{Sipe}}, \bibnamefont{and}
  \bibinfo{author}{\bibfnamefont{H.~M.} \bibnamefont{van Driel}},
  \bibinfo{journal}{Phys. Rev. B} \textbf{\bibinfo{volume}{72}},
  \bibinfo{pages}{201302(R)} (\bibinfo{year}{2005}).

\bibitem[{\citenamefont{H$\mathrm{\ddot{a}}$gele
  et~al.}(1998)\citenamefont{H$\mathrm{\ddot{a}}$gele, Oestreich,
  R$\mathrm{\ddot{u}}$hle, Nestle, and Eberl}}]{apl731580}
\bibinfo{author}{\bibfnamefont{D.}~\bibnamefont{H$\mathrm{\ddot{a}}$gele}},
  \bibinfo{author}{\bibfnamefont{M.}~\bibnamefont{Oestreich}},
  \bibinfo{author}{\bibfnamefont{W.~W.} \bibnamefont{R$\mathrm{\ddot{u}}$hle}},
  \bibinfo{author}{\bibfnamefont{N.}~\bibnamefont{Nestle}}, \bibnamefont{and}
  \bibinfo{author}{\bibfnamefont{K.}~\bibnamefont{Eberl}},
  \bibinfo{journal}{Appl. Phys. Lett.} \textbf{\bibinfo{volume}{73}},
  \bibinfo{pages}{1580} (\bibinfo{year}{1998}).

\bibitem[{\citenamefont{Kikkawa and Awschalom}(1999)}]{n397139}
\bibinfo{author}{\bibfnamefont{J.~M.} \bibnamefont{Kikkawa}} \bibnamefont{and}
  \bibinfo{author}{\bibfnamefont{D.~D.} \bibnamefont{Awschalom}},
  \bibinfo{journal}{Nature} \textbf{\bibinfo{volume}{397}},
  \bibinfo{pages}{139} (\bibinfo{year}{1999}).

\bibitem[{\citenamefont{Tsukagoshi et~al.}(1999)\citenamefont{Tsukagoshi,
  Alphenaar, and Ago}}]{n401572}
\bibinfo{author}{\bibfnamefont{K.}~\bibnamefont{Tsukagoshi}},
  \bibinfo{author}{\bibfnamefont{B.~W.} \bibnamefont{Alphenaar}},
  \bibnamefont{and} \bibinfo{author}{\bibfnamefont{H.}~\bibnamefont{Ago}},
  \bibinfo{journal}{Nature} \textbf{\bibinfo{volume}{401}},
  \bibinfo{pages}{572} (\bibinfo{year}{1999}).

\bibitem[{\citenamefont{Crooker et~al.}(2005)\citenamefont{Crooker, Furis, Lou,
  Adelmann, Smith, Palmstrom, and Crowell}}]{s3092191}
\bibinfo{author}{\bibfnamefont{S.~A.} \bibnamefont{Crooker}},
  \bibinfo{author}{\bibfnamefont{M.}~\bibnamefont{Furis}},
  \bibinfo{author}{\bibfnamefont{X.}~\bibnamefont{Lou}},
  \bibinfo{author}{\bibfnamefont{C.}~\bibnamefont{Adelmann}},
  \bibinfo{author}{\bibfnamefont{D.~L.} \bibnamefont{Smith}},
  \bibinfo{author}{\bibfnamefont{C.~J.} \bibnamefont{Palmstrom}},
  \bibnamefont{and} \bibinfo{author}{\bibfnamefont{P.~A.}
  \bibnamefont{Crowell}}, \bibinfo{journal}{Science}
  \textbf{\bibinfo{volume}{309}}, \bibinfo{pages}{2191} (\bibinfo{year}{2005}).

\bibitem[{\citenamefont{Appelbaum et~al.}(2007)\citenamefont{Appelbaum, Huang,
  and Monsma}}]{n447295}
\bibinfo{author}{\bibfnamefont{I.}~\bibnamefont{Appelbaum}},
  \bibinfo{author}{\bibfnamefont{B.~Q.} \bibnamefont{Huang}}, \bibnamefont{and}
  \bibinfo{author}{\bibfnamefont{D.~J.} \bibnamefont{Monsma}},
  \bibinfo{journal}{Nature} \textbf{\bibinfo{volume}{447}},
  \bibinfo{pages}{295} (\bibinfo{year}{2007}).

\bibitem[{\citenamefont{Mal'shukov et~al.}(2003)\citenamefont{Mal'shukov, Tang,
  Chu, and Chao}}]{b68233307}
\bibinfo{author}{\bibfnamefont{A.~G.} \bibnamefont{Mal'shukov}},
  \bibinfo{author}{\bibfnamefont{C.~S.} \bibnamefont{Tang}},
  \bibinfo{author}{\bibfnamefont{C.~S.} \bibnamefont{Chu}}, \bibnamefont{and}
  \bibinfo{author}{\bibfnamefont{K.~A.} \bibnamefont{Chao}},
  \bibinfo{journal}{Phys. Rev. B} \textbf{\bibinfo{volume}{68}},
  \bibinfo{pages}{233307} (\bibinfo{year}{2003}).

\bibitem[{\citenamefont{Rumyantsev and Sipe}(2006)}]{b73201302}
\bibinfo{author}{\bibfnamefont{I.}~\bibnamefont{Rumyantsev}} \bibnamefont{and}
  \bibinfo{author}{\bibfnamefont{J.~E.} \bibnamefont{Sipe}},
  \bibinfo{journal}{Phys. Rev. B} \textbf{\bibinfo{volume}{73}},
  \bibinfo{pages}{201302(R)} (\bibinfo{year}{2006}).

\bibitem[{\citenamefont{Stevens et~al.}(2002)\citenamefont{Stevens, Smirl,
  Bhat, Sipe, and van Driel}}]{jap914382}
\bibinfo{author}{\bibfnamefont{M.~J.} \bibnamefont{Stevens}},
  \bibinfo{author}{\bibfnamefont{A.~L.} \bibnamefont{Smirl}},
  \bibinfo{author}{\bibfnamefont{R.~D.~R.} \bibnamefont{Bhat}},
  \bibinfo{author}{\bibfnamefont{J.~E.} \bibnamefont{Sipe}}, \bibnamefont{and}
  \bibinfo{author}{\bibfnamefont{H.~M.} \bibnamefont{van Driel}},
  \bibinfo{journal}{J. Appl. Phys.} \textbf{\bibinfo{volume}{91}},
  \bibinfo{pages}{4382} (\bibinfo{year}{2002}).

\bibitem[{\citenamefont{Ganichev et~al.}(2000)\citenamefont{Ganichev, Ketterl,
  Prettl, Ivchenko, and Vorobjev}}]{apl773146}
\bibinfo{author}{\bibfnamefont{S.~D.} \bibnamefont{Ganichev}},
  \bibinfo{author}{\bibfnamefont{H.}~\bibnamefont{Ketterl}},
  \bibinfo{author}{\bibfnamefont{W.}~\bibnamefont{Prettl}},
  \bibinfo{author}{\bibfnamefont{E.~L.} \bibnamefont{Ivchenko}},
  \bibnamefont{and} \bibinfo{author}{\bibfnamefont{L.~E.}
  \bibnamefont{Vorobjev}}, \bibinfo{journal}{Appl. Phys. Lett.}
  \textbf{\bibinfo{volume}{77}}, \bibinfo{pages}{3146} (\bibinfo{year}{2000}).

\bibitem[{\citenamefont{Ganichev et~al.}(2003)\citenamefont{Ganichev, Bel'kov,
  Schneider, Ivchenko, Tarasenko, Wegscheider, Weiss, Schuh, Beregulin, and
  Prettl}}]{b68035319}
\bibinfo{author}{\bibfnamefont{S.~D.} \bibnamefont{Ganichev}},
  \bibinfo{author}{\bibfnamefont{V.~V.} \bibnamefont{Bel'kov}},
  \bibinfo{author}{\bibfnamefont{P.}~\bibnamefont{Schneider}},
  \bibinfo{author}{\bibfnamefont{E.~L.} \bibnamefont{Ivchenko}},
  \bibinfo{author}{\bibfnamefont{S.~A.} \bibnamefont{Tarasenko}},
  \bibinfo{author}{\bibfnamefont{W.}~\bibnamefont{Wegscheider}},
  \bibinfo{author}{\bibfnamefont{D.}~\bibnamefont{Weiss}},
  \bibinfo{author}{\bibfnamefont{D.}~\bibnamefont{Schuh}},
  \bibinfo{author}{\bibfnamefont{E.~V.} \bibnamefont{Beregulin}},
  \bibnamefont{and} \bibinfo{author}{\bibfnamefont{W.}~\bibnamefont{Prettl}},
  \bibinfo{journal}{Phys. Rev. B} \textbf{\bibinfo{volume}{68}},
  \bibinfo{pages}{035319} (\bibinfo{year}{2003}).

\bibitem[{\citenamefont{Diehl et~al.}(2007)\citenamefont{Diehl, Shalygin,
  Bel'kov, Hoffmann, Danilov, Herrle, Tarasenko, Schuh, Gerl, Wegscheider
  et~al.}}]{njp9349}
\bibinfo{author}{\bibfnamefont{H.}~\bibnamefont{Diehl}},
  \bibinfo{author}{\bibfnamefont{V.~A.} \bibnamefont{Shalygin}},
  \bibinfo{author}{\bibfnamefont{V.~V.} \bibnamefont{Bel'kov}},
  \bibinfo{author}{\bibfnamefont{C.}~\bibnamefont{Hoffmann}},
  \bibinfo{author}{\bibfnamefont{S.~N.} \bibnamefont{Danilov}},
  \bibinfo{author}{\bibfnamefont{T.}~\bibnamefont{Herrle}},
  \bibinfo{author}{\bibfnamefont{S.~A.} \bibnamefont{Tarasenko}},
  \bibinfo{author}{\bibfnamefont{D.}~\bibnamefont{Schuh}},
  \bibinfo{author}{\bibfnamefont{C.}~\bibnamefont{Gerl}},
  \bibinfo{author}{\bibfnamefont{W.}~\bibnamefont{Wegscheider}},
  \bibnamefont{et~al.}, \bibinfo{journal}{New J. Phys.}
  \textbf{\bibinfo{volume}{9}}, \bibinfo{pages}{349} (\bibinfo{year}{2007}).

\bibitem[{\citenamefont{Ganichev et~al.}(2001)\citenamefont{Ganichev, Ivchenko,
  Danilov, Eroms, Wegscheider, Weiss, and Prettl}}]{l864358}
\bibinfo{author}{\bibfnamefont{S.~D.} \bibnamefont{Ganichev}},
  \bibinfo{author}{\bibfnamefont{E.~L.} \bibnamefont{Ivchenko}},
  \bibinfo{author}{\bibfnamefont{S.~N.} \bibnamefont{Danilov}},
  \bibinfo{author}{\bibfnamefont{J.}~\bibnamefont{Eroms}},
  \bibinfo{author}{\bibfnamefont{W.}~\bibnamefont{Wegscheider}},
  \bibinfo{author}{\bibfnamefont{D.}~\bibnamefont{Weiss}}, \bibnamefont{and}
  \bibinfo{author}{\bibfnamefont{W.}~\bibnamefont{Prettl}},
  \bibinfo{journal}{Phys. Rev. Lett.} \textbf{\bibinfo{volume}{86}},
  \bibinfo{pages}{4358} (\bibinfo{year}{2001}).

\bibitem[{\citenamefont{Wei et~al.}(2007)\citenamefont{Wei, Cho, Chen, Peng,
  Chiu, and Kuan}}]{apl91252102}
\bibinfo{author}{\bibfnamefont{C.~M.} \bibnamefont{Wei}},
  \bibinfo{author}{\bibfnamefont{K.~S.} \bibnamefont{Cho}},
  \bibinfo{author}{\bibfnamefont{Y.~F.} \bibnamefont{Chen}},
  \bibinfo{author}{\bibfnamefont{Y.~H.} \bibnamefont{Peng}},
  \bibinfo{author}{\bibfnamefont{C.~W.} \bibnamefont{Chiu}}, \bibnamefont{and}
  \bibinfo{author}{\bibfnamefont{C.~H.} \bibnamefont{Kuan}},
  \bibinfo{journal}{Appl. Phys. Lett.} \textbf{\bibinfo{volume}{91}},
  \bibinfo{pages}{252102} (\bibinfo{year}{2007}).

\bibitem[{\citenamefont{Cho et~al.}(2007{\natexlab{a}})\citenamefont{Cho,
  Liang, Chen, Tang, and Shen}}]{b75085327}
\bibinfo{author}{\bibfnamefont{K.~S.} \bibnamefont{Cho}},
  \bibinfo{author}{\bibfnamefont{C.~T.} \bibnamefont{Liang}},
  \bibinfo{author}{\bibfnamefont{Y.~F.} \bibnamefont{Chen}},
  \bibinfo{author}{\bibfnamefont{Y.~Q.} \bibnamefont{Tang}}, \bibnamefont{and}
  \bibinfo{author}{\bibfnamefont{B.}~\bibnamefont{Shen}},
  \bibinfo{journal}{Phys. Rev. B} \textbf{\bibinfo{volume}{75}},
  \bibinfo{pages}{085327} (\bibinfo{year}{2007}{\natexlab{a}}).

\bibitem[{\citenamefont{Cho et~al.}(2007{\natexlab{b}})\citenamefont{Cho, Chen,
  Tang, and Shen}}]{apl90041909}
\bibinfo{author}{\bibfnamefont{K.~S.} \bibnamefont{Cho}},
  \bibinfo{author}{\bibfnamefont{Y.~F.} \bibnamefont{Chen}},
  \bibinfo{author}{\bibfnamefont{Y.~Q.} \bibnamefont{Tang}}, \bibnamefont{and}
  \bibinfo{author}{\bibfnamefont{B.}~\bibnamefont{Shen}},
  \bibinfo{journal}{Appl. Phys. Lett.} \textbf{\bibinfo{volume}{90}},
  \bibinfo{pages}{041909} (\bibinfo{year}{2007}{\natexlab{b}}).

\bibitem[{\citenamefont{He et~al.}(2007)\citenamefont{He, Shen, Tang, Tang,
  Yin, Xu, Yang, Zhang, Chen, Tang et~al.}}]{apl91071912}
\bibinfo{author}{\bibfnamefont{X.~W.} \bibnamefont{He}},
  \bibinfo{author}{\bibfnamefont{B.}~\bibnamefont{Shen}},
  \bibinfo{author}{\bibfnamefont{Y.~Q.} \bibnamefont{Tang}},
  \bibinfo{author}{\bibfnamefont{N.}~\bibnamefont{Tang}},
  \bibinfo{author}{\bibfnamefont{C.~M.} \bibnamefont{Yin}},
  \bibinfo{author}{\bibfnamefont{F.~J.} \bibnamefont{Xu}},
  \bibinfo{author}{\bibfnamefont{Z.~J.} \bibnamefont{Yang}},
  \bibinfo{author}{\bibfnamefont{G.~Y.} \bibnamefont{Zhang}},
  \bibinfo{author}{\bibfnamefont{Y.~H.} \bibnamefont{Chen}},
  \bibinfo{author}{\bibfnamefont{C.~G.} \bibnamefont{Tang}},
  \bibnamefont{et~al.}, \bibinfo{journal}{Appl. Phys. Lett.}
  \textbf{\bibinfo{volume}{91}}, \bibinfo{pages}{071912}
  (\bibinfo{year}{2007}).

\bibitem[{\citenamefont{Tang et~al.}(2007)\citenamefont{Tang, Shen, He, Han,
  Tang, Chen, Yang, Zhang, Chen, Tang et~al.}}]{apl91071920}
\bibinfo{author}{\bibfnamefont{Y.~Q.} \bibnamefont{Tang}},
  \bibinfo{author}{\bibfnamefont{B.}~\bibnamefont{Shen}},
  \bibinfo{author}{\bibfnamefont{X.~W.} \bibnamefont{He}},
  \bibinfo{author}{\bibfnamefont{K.}~\bibnamefont{Han}},
  \bibinfo{author}{\bibfnamefont{N.}~\bibnamefont{Tang}},
  \bibinfo{author}{\bibfnamefont{W.~H.} \bibnamefont{Chen}},
  \bibinfo{author}{\bibfnamefont{Z.~J.} \bibnamefont{Yang}},
  \bibinfo{author}{\bibfnamefont{G.~Y.} \bibnamefont{Zhang}},
  \bibinfo{author}{\bibfnamefont{Y.~H.} \bibnamefont{Chen}},
  \bibinfo{author}{\bibfnamefont{C.~G.} \bibnamefont{Tang}},
  \bibnamefont{et~al.}, \bibinfo{journal}{Appl. Phys. Lett.}
  \textbf{\bibinfo{volume}{91}}, \bibinfo{pages}{071920}
  (\bibinfo{year}{2007}).

\bibitem[{\citenamefont{Bhat and Sipe}(2000)}]{l855432}
\bibinfo{author}{\bibfnamefont{R.~D.~R.} \bibnamefont{Bhat}} \bibnamefont{and}
  \bibinfo{author}{\bibfnamefont{J.~E.} \bibnamefont{Sipe}},
  \bibinfo{journal}{Phys. Rev. Lett.} \textbf{\bibinfo{volume}{85}},
  \bibinfo{pages}{5432} (\bibinfo{year}{2000}).

\bibitem[{\citenamefont{Najmaie et~al.}(2003)\citenamefont{Najmaie, Bhat, and
  Sipe}}]{b68165348}
\bibinfo{author}{\bibfnamefont{A.}~\bibnamefont{Najmaie}},
  \bibinfo{author}{\bibfnamefont{R.~D.~R.} \bibnamefont{Bhat}},
  \bibnamefont{and} \bibinfo{author}{\bibfnamefont{J.~E.} \bibnamefont{Sipe}},
  \bibinfo{journal}{Phys. Rev. B} \textbf{\bibinfo{volume}{68}},
  \bibinfo{pages}{165348} (\bibinfo{year}{2003}).

\bibitem[{\citenamefont{Meier and Zakharchenya}(1984)}]{opticalorientation}
\bibinfo{author}{\bibfnamefont{F.}~\bibnamefont{Meier}} \bibnamefont{and}
  \bibinfo{author}{\bibfnamefont{B.~P.} \bibnamefont{Zakharchenya}},
  \emph{\bibinfo{title}{Optical orientation}}, vol.~\bibinfo{volume}{8} of
  \emph{\bibinfo{series}{Modern Problems in Condensed Matter Sciences}}
  (\bibinfo{publisher}{North-Holland, Amsterdam}, \bibinfo{year}{1984}).

\bibitem[{\citenamefont{Zhao}(2008)}]{apl92112104}
\bibinfo{author}{\bibfnamefont{H.}~\bibnamefont{Zhao}}, \bibinfo{journal}{Appl.
  Phys. Lett.} \textbf{\bibinfo{volume}{92}}, \bibinfo{pages}{112104}
  (\bibinfo{year}{2008}).

\bibitem[{\citenamefont{Schmitt-Rink et~al.}(1985)\citenamefont{Schmitt-Rink,
  Chemla, and Miller}}]{b326601}
\bibinfo{author}{\bibfnamefont{S.}~\bibnamefont{Schmitt-Rink}},
  \bibinfo{author}{\bibfnamefont{D.~S.} \bibnamefont{Chemla}},
  \bibnamefont{and} \bibinfo{author}{\bibfnamefont{D.~A.~B.}
  \bibnamefont{Miller}}, \bibinfo{journal}{Phys. Rev. B}
  \textbf{\bibinfo{volume}{32}}, \bibinfo{pages}{6601} (\bibinfo{year}{1985}).

\end{thebibliography}
\end{document}